\documentclass[aps,twocolumn,superscriptaddress,prd,nofootinbib]{revtex4}

\usepackage{graphicx}
\usepackage{url}
\usepackage{multirow}
\usepackage{amsmath}
\usepackage{amssymb}
\usepackage{amsfonts}
\usepackage[colorlinks,linkcolor=blue,anchorcolor=blue,citecolor=red]{hyperref}
\usepackage{xcolor}
\usepackage{acronym}
\usepackage{ulem} 
\usepackage{CJK}
\usepackage{orcidlink}

\newcommand{\TRC}{MOE Key Laboratory of TianQin Mission, TianQin Research Center for Gravitational Physics \& School of Physics and Astronomy, Frontiers Science Center for TianQin, Gravitational Wave Research Center of CNSA, Sun Yat-sen University (Zhuhai Campus), Zhuhai 519082, China}
\newcommand{\UoP}{University of Portsmouth, Portsmouth, PO1 3FX, United Kingdom}

\begin{document}
\begin{CJK*}{UTF8}{gbsn}

\title{Archival Inference for Eccentric Stellar-Mass Binary Black Holes in Space-Based Gravitational Wave Observations}

\author{Han Wang(王晗)\orcidlink{0009-0007-5095-9227}}
\affiliation{Kavli Institute for Astronomy and Astrophysics, Peking University, Beijing 100871, China}
\affiliation{\TRC}
\affiliation{\UoP}

\author{Michael J. Williams\orcidlink{0000-0003-2198-2974}}
\affiliation{\UoP}

\author{Ian Harry\orcidlink{0000-0002-5304-9372}}
\affiliation{\UoP}

\author{Yi-Ming Hu(胡一鸣)\orcidlink{0000-0002-7869-0174}}
\email{huyiming@mail.sysu.edu.cn}
\affiliation{\TRC}

\date{\today}

\begin{abstract}
	Space-based gravitational-wave observatories will detect the early inspiral of stellar-mass binary black holes and can track their eccentricity evolution.
	However, untargeted searches in the space band are computationally demanding and require relatively high detection thresholds (signal-to-noise ratio $\sim 15$).
	Information from ground-based detections can significantly shrink the parameter space for space-band analyses and thereby substantially reduce the detection threshold.
	We present a Bayesian inference pipeline for ground-triggered archival space-band analyses that includes eccentricity.
	Using ground-informed priors, we demonstrate that with one year of LISA or TianQin data a GW190521-like source with signal-to-noise ratio $\sim 7$ can be distinguished and tightly constrained.
	In this setup, space observations sharpened the redshifted chirp mass from $\mathcal{O}(10^{-3})M_\odot$ to $\mathcal{O}(10^{-5})M_\odot$, and constrain the eccentricity to $\mathcal{O}(10^{-5})$ around the injected value $e_{0.01\mathrm{Hz}}=0.1$.
	These results demonstrate that inference of eccentric stellar-mass binary black holes in noisy space-band data is practically feasible, supports an expanded yield of multiband detections, and strengthens prospects for future astrophysical and gravitational tests.
\end{abstract}

\maketitle
\end{CJK*}

\acrodef{GW}{gravitational wave}
\acrodef{sBBH}{stellar-mass binary black hole}
\acrodef{EM}{electromagnetic}
\acrodef{SNR}{signal-to-noise ratio}
\acrodef{FIM}{Fisher information matrix}
\acrodef{MCMC}{Markov Chain Monte Carlo}
\acrodef{PSD}{power spectral density}
\acrodef{LISA}{Laser Interferometer Space Antenna}
\acrodef{PN}{post Newtonian}
\acrodef{TDI}{Time delay interferometry}
\acrodef{SPA}{stationary phase approximation}
\acrodef{AGN}{active galactic nuclei}

\section{Introduction}
Since the first detection of \acp{GW} by LIGO and Virgo in 2015 \cite{Abbott2016}, hundreds of \ac{GW} events have been reported, the majority of which are \ac{sBBH} mergers with component masses in the tens of solar masses \cite{Abbott2023a,Abac2025a,Abac2025b}.
The ever-expanding catalog of \ac{GW} detections promises to enrich our understanding of the physical characteristics of \acp{sBBH}\cite{Abbott2020,Abbott2020c,Abbott2021a,Abbott2023,Abac2025b}.
The orbital eccentricity of an \ac{sBBH} system is a key probe to study its formation channel.
If \acp{sBBH} are formed by a co-evolution process from two massive stars, their orbits are expected to be almost circular, while \acp{sBBH} formed by dynamical interaction are usually expected to be born with non-negligible eccentricities\cite{Belczynski2016,Barack2019,Breivik2016,Rodriguez2016}.

It is possible for a system in the ground-based observation window to retain a measurable eccentricity if it was formed by dynamical interactions very close to merger, and investigations of these events aiming at constraining the eccentricity have been performed \cite{Abbott2019,RomeroShaw2019}.
GW190521 is an event with a total mass of $ \sim 150 M_\odot $ and is the first event argued to have significant orbital eccentricity (eccentricity at 10 Hz $ e_{10} \gtrsim 0.1 $) \cite{Abbott2020c, RomeroShaw2020}. Some studies suggest that its eccentricity $ e_{10}$ may be as large as $ \sim 0.7 $ \cite{Gayathri2022}.
An additional three events have shown hints of eccentricity ($ e_{10}\geq0.1 $) after GW190521, providing more evidence for the dynamical formation channel \cite{RomeroShaw2021,RomeroShaw2022}.
However, current ground-based detectors are only sensitive to \acp{sBBH} in the final seconds before coalescence due to their limited frequency band.
By that time, the systems have typically circularized, resulting in a lack of strong evidence to confirm whether these events originated from eccentric mergers.

In the near future, space-based \ac{GW} observatories, such as TianQin\cite{Luo2016} and \ac{LISA}\cite{AmaroSeoane2017}, are expected to be launched and observe the universe. 
Space detectors operate in a lower-frequency sensitivity band due to their longer baselines, enabling them to observe \acp{sBBH} over much longer timescales during their inspiral phase.
Years of observation allow space detectors to track the evolution of key physical characteristics, such as eccentricity and spin, and significantly enhance the precision in measuring the source masses \cite{Nishizawa2016,Liu2020,Klein2022,Wang2024}.
Even when highly eccentric systems fall outside the space band at formation\cite{Chen2017,Samsing2018}, the resulting non-detection can help distinguish high-eccentricity formation mechanisms via upper limits on space-band \ac{GW} emission.

Data analysis for space-based detectors, usually divided into two stages -- detection and parameter estimation -- presents multiple challenges.
The detection stage focuses mostly on determining the existence of astrophysical signals in the data, where the matched-filtering method is widely used~\cite{Allen:2005fk}.
A matched filtering pipeline typically requires constructing a ``template bank'' of \ac{GW} waveforms covering the relevant \ac{GW} parameter space~\cite{Owen:1998dk}.
However, space-based searches for \acp{sBBH} require impractically large template banks due to the long signal durations.
\citet{Moore2019a} predicts that covering the full \ac{sBBH} space for LISA needs $\mathcal{O}(10^{30})$ templates.
Since a much larger template bank implies many more independent trials on the noise, maintaining a fixed false-alarm probability requires raising the effective detection threshold to about $\rho_{\rm thr}\sim15$ (versus 8 for LIGO/Virgo).
Combined with the slower \ac{SNR} accumulation at lower frequencies for \acp{sBBH}, this means that far fewer \acp{sBBH} would be loud enough to detect.
For example, even assuming a uniform-in-volume population, increasing the threshold from 8 to 15 reduces the sensitive volume to $\simeq0.15$ ($\approx 6 \times$ fewer detections), and the actual reduction is likely larger given the realistic merger rate \cite{Abac2025b}.
The same issues also affect the parameter estimation phase, and further raise the computational burden.
Some studies have investigated parameter estimation for \acp{sBBH}  under the noise-free assumption \cite{Toubiana2020,Buscicchio2021,Lyu2023}.
Since practical inference must incorporate detector noise, developing a viable end-to-end analysis framework for \acp{sBBH} remains an open question.

Several strategies have been proposed to mitigate these challenges.
Semi-coherent searches, which split the data into segments and match each separately, combined with particle swarm optimization might make \ac{sBBH} searches tractable~\cite{Bandopadhyay2023,Bandopadhyay2024,Bandopadhyay2025,Fu2025}.
Additionally, machine-learning-based detection methods have also been explored~\cite{Zhang2024}.
For parameter estimation, a wavelet-based likelihood can speed up Bayesian inference by orders of magnitude over standard frequency-domain approaches and enables possible treatment of non-stationary noise \cite{Digman2023}.
Despite these advances, the fundamental requirement of requiring \ac{SNR} $(\gtrsim15)$ to detect \ac{sBBH} signals remains.

Multiband joint detection offers a promising way to observe additional \acp{sBBH} with space-based observatories~\cite{Sesana2016,Klein2022,Toubiana2022}.
Next-generation ground-based detectors such as the Einstein Telescope (ET) \cite{Punturo2010} and Cosmic Explorer (CE) \cite{Evans2023} will detect $\mathcal{O}(10^5)$ events per year\cite{Abac2025c} and identify events with most source parameters tightly constrained.
That information can then be used to revisit space-based data archives and recover the signal from the same source (so-called ``archival searches'') \cite{Wong2018,Ewing2021}, thereby restricting the region of parameter space to be searched and significantly reducing the computational burden.
In \citet{Wang2024}, we implemented a matched-filtering bank generation pipeline for such archival searches with eccentricity included.
Multiband data also enable coherent parameter estimation.
By marginalizing over extrinsic parameters, efficient inference remains possible even at LISA \ac{SNR} $\sim3$ \cite{Wu2025}.

In this paper, following \citet{Wang2024}, we develop a Bayesian inference pipeline for \ac{sBBH} signals from archival searches that allows for orbital eccentricity.
Section~\ref{sec2} details the methodology, including the Bayesian-inference configuration and an eccentricity-compatible frequency-domain antenna response for space-based detectors.
Section~\ref{sec3} quantifies the parameter-estimation precision achievable with next-generation ground detectors for a GW190521-like source, showing that the space-based follow-up need only search over the masses and eccentricity.
Section~\ref{sec4} presents results of space-based Bayesian parameter estimation.
Section~\ref{sec5} summarizes the results and discusses implications.
We will use units $ G=c=1 $ unless otherwise specified.

\section{Methodology\label{sec2}}

In this section, we introduce the methods and configurations used in our Bayesian inference framework for \ac{GW} parameter estimation.
The posterior of a parameter set $\lambda^\mu$ given data $D$ follows from Bayes' theorem:
\begin{equation}
	P(\lambda^\mu\mid D)
	= \frac{P(D\mid \lambda^\mu)P(\lambda^\mu)}{P(D)}
	\propto\mathcal{L}(\lambda^\mu)P(\lambda^\mu),
\end{equation}
where $P(\lambda^\mu)$ encodes prior information, $\mathcal{L}\equiv P(D\mid \lambda^\mu)$ is the likelihood, and Bayesian evidence
\begin{equation}\label{evidence}
	P(D)=\int P(D\mid\lambda^\mu)P(\lambda^\mu)\mathrm{d}\lambda^\mu
\end{equation}
serves only to normalize the posterior.
Under the usual assumption of stationary, Gaussian detector noise $n(t)$, with $D(t)=h(t,\lambda^\mu)+n(t)$, the logarithm of the likelihood can be written as
\begin{equation}\label{likelihood}
	\begin{aligned}
		\ln \mathcal{L}&\propto -\frac{1}{2}(D-h(\lambda^\mu)\mid D-h(\lambda^\mu))\\
		&=-\frac{1}{2}\left[(D\mid D)+(h\mid h)-2(h\mid D)\right],
	\end{aligned}
\end{equation}
using the inner product
\begin{equation}
	(h \mid g)\equiv 4 \Re \int_0^{\infty} \frac{\tilde{g}^*(f) \tilde{h}(f)}{S_n(f)} \mathrm{d} f,
\end{equation}
where $\tilde h(f)$ is the Fourier transform of the waveform and $S_n(f)$ is the one-sided detector noise \ac{PSD}.
To accelerate likelihood evaluations for long, slowly-varying signals, we employ a heterodyned (or ``relative-binning'') approximation:
a single high-resolution reference waveform is computed once, and subsequent waveforms are recovered by interpolating their ratio to that reference over coarse frequency bins \cite{Cornish2010,Zackay2018,Leslie2021}.
We also employ \texttt{nessai}, a nested-sampling algorithm augmented with normalizing flows, to accelerate sampling efficiency\cite{Williams2021,Williams2023}.
All analyses are performed within the \texttt{PyCBC Inference} framework \cite{Biwer2019}.

We assume that signals in ground-based detectors have negligible eccentricity and, for consistency with the non-spinning waveform models used below in space-based analysis, we also neglect component spins.
Thus, we adopt a non-spinning, circular parameter set
$$ \lambda^\mu_\mathrm{g} = \left(\mathcal{M},q,D_L,t_c,\phi_c,\iota,\alpha,\delta,\psi \right), $$
where 
$$ \mathcal{M}=(m_1m_2)^{3/5}(m_1+m_2)^{-1/5},\quad q \equiv \frac{m_1}{m_2}\ge1, $$
with $m_1$ and $m_2$ being the component masses,
$D_L$ is the luminosity distance, $(t_c,\phi_c)$ the coalescence time and phase, $\iota$ the inclination, $(\alpha,\delta)$ the right-ascension and declination in equatorial coordinates, and $\psi$ the polarization.
For the space-based analysis, we add eccentricity and switch to ecliptic sky coordinates:
$$ \lambda^\mu_\mathrm{s} = \left(\mathcal{M},q,D_L,t_c,\phi_c,\iota,\lambda,\beta,\psi,e_{0.01\mathrm{Hz}} \right), $$
where $(\lambda,\beta)$ are the ecliptic longitude and latitude in the solar-system barycenter (SSB) frame, and $e_{0.01\mathrm{Hz}}$ is the eccentricity at the reference GW frequency $f_{\rm ref}=0.01 \mathrm{Hz}$, a sweet spot that lies in the mHz band where space-based detectors are most sensitive and where \acp{sBBH} typically reside years before merger.

\subsection{Detector configurations}
We consider a third-generation ground network composed of ET and two CE detectors.
ET, assumed here to be located in Italy, adopts a triangular configuration with three co-located interferometers with 10km armlength \cite{Punturo2010}, with the sensitivity, location, and orientation defined in \cite{LSC2018}.
The two CE detectors are placed off the coasts of Washington state and Texas, with 40km and 20km arms, respectively \cite{Evans2023}.
Their full location and orientation are defined in \cite{Wu2025,multiband}.
\footnote{We note that these locations are used for simulation purposes only and do not represent the final sites of these instruments, which are still under active discussion.}
Note that these configurations are only for simulation and do not represent the real plan of these instruments.
For ground analyses we employ the frequency-domain waveform model \texttt{IMRPhenomXPHM}\cite{Pratten2021} for both signal injection and recovery, which models quasi-circular, precessing binary black holes and includes higher multipoles.
The low-frequency cutoffs are set to $ f_\mathrm{low}=2 \mathrm{Hz} $ for ET and $ f_\mathrm{low}=5 \mathrm{Hz} $ for CE.

For space-based observations we consider both single-detector (TianQin or LISA) and joint TianQin+LISA network configurations.
The inspiral of eccentric binaries is modeled with the non-spinning frequency-domain \texttt{EccentricFD} approximation\cite{Yunes2009,Huerta2014}.
We assume a continuous one-year observation preceding coalescence and neglect duty cycles for simplicity.
Unlike ground detectors with fixed armlengths, space constellations have time-varying, generally unequal arms due to spacecraft motion.
\ac{TDI} is therefore used to cancel laser phase noise by forming specific time-delayed combinations.
In this work, we analyze the orthogonal TDI channels $(A,E,T)$ (details in the next subsection).
For TianQin and LISA, the residual acceleration and position measurement noise levels follow \cite{Luo2016} and \cite{Babak2021}, respectively.
We adopt detector orbits derived by \citet{Hu2018} for TianQin, and \citet{Cornish2003} for LISA.

We emphasize that different waveform families are used in the ground and space bands, and these models are not fully consistent.
This choice reflects suitability and computational efficiency for archival analyses.
A unified approximant that contains both spin and eccentricity across both bands will be adopted once available.

\subsection{Antenna response for eccentric signals}

In this subsection, we briefly review the antenna responses for ground and space detectors and the eccentric-harmonic structure of the \texttt{EccentricFD} waveform.
We then present the edited frequency-domain \ac{TDI} response tailored to eccentric signals, which was used in \cite{Wang2024} and is documented here in detail.

In the transverse-traceless gauge, a \ac{GW} signal described by two polarizations ($ h_+ $ and $ h_\times $) can be written as
\begin{equation}\label{Hwf}
	H=h_{+} P_{+}+h_{ \times} P_{ \times},
\end{equation}
where $P_{+}$ and $P_{\times}$ are polarization tensors.
Let $(u,v)$ be orthogonal basis vectors transverse to the propagation direction $k$, for zero polarization angle,
\begin{equation}
	\begin{aligned}
		P_{+}^0(\lambda,\beta)=P_{+}(\psi=0)=u \otimes u-v\otimes v ,\\
		P_{\times}^0(\lambda,\beta)=P_{\times}(\psi=0)=u\otimes v+v \otimes u ,
	\end{aligned}
\end{equation}
where $\otimes$ denotes the dyadic (outer) product: $ (a\otimes b)_{ij} = a_i b_j$.
Then a rotation by the polarization angle $\psi$ gives
\begin{equation}
	P_{+}+i P_{\times}=e^{-2 i \psi}\left(P_{+}^{0}+i P_{\times}^{0}\right).
\end{equation}

Ground detectors observe the \ac{GW} signal only for minutes, if not seconds, before coalescence for the multiband sources of interest here, which are more likely to be high-mass \acp{sBBH}.
The Doppler frequency modulation from the Earth's orbital motion can be neglected, and $(P_{+},P_{\times})$ may be treated as constant for a given sky location.
In contrast, the long observation time and orbital motion of space detectors make the response time dependent.
This is further compounded for eccentric binaries, which radiate in multiple harmonics.

\subsubsection{Eccentric harmonics}
\ac{GW} radiation drives binaries to circularize over time\cite{Peters1964}.
At leading \ac{PN} order the eccentricity scales as
$e \sim e_0 (f/f_0)^{-19/18}$\cite{Yunes2009}.
For example, a system with initial eccentricity $e_0=0.1$ at initial frequency $f_0=0.01\mathrm{Hz}$ will enter the ground band ($f\gtrsim1\mathrm{Hz}$) with $e\lesssim10^{-3}$.
Thus sources that appear circular on the ground can still be measurably eccentric in the space band.

We model space-band inspirals with the nonspinning, frequency-domain \texttt{EccentricFD} waveform~\cite{Yunes2009,Huerta2014}, an inspiral-only approximation valid for $e_0\lesssim0.4$.
Because \acp{sBBH} typically merge above the sensitive frequency band of space-based detectors, this model is adequate in TianQin or LISA with moderate eccentricities.

For quasi-circular binaries, one can decompose the waveform in spin-weighted spherical harmonics as
\begin{equation}
	h_{+}-i h_{\times}=\sum_{\ell \geq 2} \sum_{m=-\ell}^{\ell}{ }_{-2} Y_{\ell m}(\iota, \varphi) h_{\ell m},
\end{equation}
where $ \iota $ and $ \varphi $ represent the direction of \ac{GW} emission in the source frame.
The dominant harmonic is $ (\ell, m)=(2, 2) $ and other modes are referred to as higher-order modes.
Substituting into Eq. \eqref{Hwf} gives
\begin{equation}\label{hlm}
	H =\sum_{\ell, m} P_{\ell m}(\lambda,\beta,\psi,\iota,\varphi) h_{\ell m}.
\end{equation}
This is a convenient form for response calculation, as $ P_{\ell m} $ is constant for a specific signal within each harmonic\cite{Marsat2021}.

Eccentric binaries produce additional orbital harmonics that play a role similar to higher-order modes but are described by the mean orbital frequency $ F $.
Under the stationary-phase approximation,
\begin{equation}
	f = j \cdot F(t_0),
\end{equation}
where $ t_0 $ is the time which gives the stationary point of $ F $ and the dominant eccentric harmonic is $ j=2 $.
Here we use $ j $ to label eccentric harmonics (to be distinguished from the spin-weighted spherical-harmonic indices $(\ell,m)$ above).

The \texttt{EccentricFD} approximation takes the form \cite{Yunes2009}
\begin{equation}
	\tilde{h}_{+,\times}=\tilde{\mathcal{A}} f^{-7 / 6} \sum_{j=1}^{10} \xi_{j}^{+,\times}\left(\frac{j}{2}\right)^{2 / 3} e^{{\rm i}\Psi_{j}},
\end{equation}
with
\begin{equation}
	\tilde{\mathcal{A}} =  - {\left( {\frac{5}{{384}}} \right)^{1/2}}{\pi ^{ - 2/3}}\frac{{{\mathcal{M}^{5/6}}}}{{{D_L}}},
\end{equation}
and
\begin{equation}
	\xi _j^{ + , \times } = C_{ + , \times }^{(j)} + {\rm i}S_{ + , \times }^{(j)},
\end{equation}
which depend on $ (\iota, \varphi) $ and the evolving eccentricity $ e(F) $.
When $ e=0 $,
\begin{equation}
	\begin{aligned}
		\xi _{j = 2}^ +  &= C_ + ^{(2)} + {\rm i}S_ + ^{(2)} = 4 \cdot \frac{{1 + {{\cos }^2}\iota }}{2}{e^{{\rm i} \cdot 2\varphi }}, \\
		\xi _{j = 2}^ \times  &= C_ \times ^{(2)} + {\rm i}S_ \times ^{(2)} = 4 \cdot \left( { - \cos \iota } \right){e^{{\rm i} \cdot 2\varphi }}, \\
		\xi _{j \ne 2}^{ + , \times } &= 0,
	\end{aligned}
\end{equation}
which recovers the usual quadrupole coefficients.
The Fourier phase
\begin{equation}
	{\Psi _j} = \frac{\pi }{4} + j{\phi _c} - 2\pi f{t_c} + {\left( {\frac{j}{2}} \right)^{8/3}}{\Psi _{\text{PN}}^{ecc}},
\end{equation}
where $ \Psi _{\text{PN}}^{ecc} $ is the \ac{PN} phase with eccentricity \cite{Huerta2014}.
Following \texttt{TaylorF2}, the phase includes \ac{PN} corrections up to 3.5PN order.
Both the amplitude and phase are expanded to $ \mathcal{O}(e^8) $ and then re-expanded in $ e_0 $ up to $ \mathcal{O}(e_0^8) $.

We apply a bandpass so that the signal remains within each detector's valid band.
Because eccentric harmonics map differently to the Fourier frequency, we gate the waveform as
\begin{equation}
	\tilde h_\mathrm{det} = \tilde h \times \Theta \left( j \cdot f_\text{high} - 2f \right)\Theta \left( 2f - j \cdot f_\text{low} \right),
\end{equation}
where $ \Theta $ is the Heaviside function.
We also standardize the eccentricity definition by fixing a reference GW frequency $f_{\rm ref}=0.01 \mathrm{Hz}$ and parametrizing $e_{\rm ref}\equiv e(f_{\rm ref})$.
The \texttt{EccentricFD} evolution is initialized at $f_{\rm ref}$ and then band-limited by the gate above.
This ensures that all runs use a consistent eccentricity definition $e_{0.01\mathrm{Hz}}$ while preserving the correct bandpass per harmonic.

For the lower cutoff we use
$$
f_\text{low} = \max \left[ 10^{-4}\mathrm{Hz},f_0 \right],
$$
where the leading-PN start frequency can be written as
\begin{equation}
	f_0=(5 / 256)^{3 / 8} {\pi}^{-1} \mathcal{M}^{-5 / 8}T^{-3 / 8},
\end{equation}
with a one-year observation ($T=1\mathrm{yr}$).
The eccentricity corrections to $ f_0 $ are negligible for $ M_\mathrm{tot} \lesssim 10^5 M_\odot $ and $ T \gtrsim 1 \mathrm{yr} $ \cite[App.~E]{Yunes2009}.
The upper cutoff differs by mission:
$$
f_\mathrm{high}^{\rm TianQin} = \min \left[ f_\text{ISCO},1\mathrm{Hz} \right],
f_\mathrm{high}^{\rm LISA} = \min \left[ f_\text{ISCO},0.1\mathrm{Hz} \right],
$$
where $ f_\text{ISCO} = 1/(6^{3/2} \pi M_\mathrm{tot}) $, $ M_\mathrm{tot} = m_1+m_2 $.

\subsubsection{Frequency-domain \ac{TDI} response}
In the Michelson construction, the three \ac{TDI} channels (X,Y,Z) are linear combinations of single-link responses $ \tilde{y}_{sr} $ with delay operator $ z\equiv\exp(i 2\pi f L) $ (armlength $L$)\cite{Babak2021}:
\begin{equation}
	\tilde{X} = (1-z^2) \left(\tilde{y}_{31} + z \tilde{y}_{13} - \tilde{y}_{21} -z \tilde{y}_{12}\right),
\end{equation}
and $\tilde{Y}, \tilde{Z}$ follow by cyclic permutation $(1\rightarrow2\rightarrow3\rightarrow1)$.
The orthogonal channels $ (A,E,T) $ are defined by
\begin{equation}
	\begin{aligned}
		A &= \frac{1}{{\sqrt 2 }}(Z - X),\\
		E &= \frac{1}{{\sqrt 6 }}(X - 2Y + Z),\\
		T &= \frac{1}{{\sqrt 3 }}(X + Y + Z),		
	\end{aligned}
\end{equation}
whose instrumental noises can be treated as independent.
We adopt the channel \acp{PSD} given in \cite{Babak2021}.

\begin{table*}
	\centering
	\caption{Priors and injected values for a GW190521-like source.
	To ensure sufficient \ac{SNR} in the space-based data, the source is placed at a smaller luminosity distance.
	Mass parameters are redshifted under the Planck~2015 cosmology~\cite{Ade2016}. For the space analysis, equatorial coordinates $(\alpha,\delta)$ are converted to ecliptic longitude and latitude $(\lambda,\beta)$ in the SSB frame.
	Space-based priors are constructed from a KDE of the corresponding ground-based posterior (see text).}
	\begin{tabular}{c|c|c|c|c}
		\hline
		\hline
		& \multirow{2}{*}{Injected values} & \multicolumn{3}{c}{Prior} \\
		\cline{3-5}                   &                   &          ET + dual CE               &   Space:10D   & Space:3D              \\
		\hline   
		\hline  
		$ \mathcal{M} $ [$ M_\odot $] &      77.6748      &            Uniform(77, 79)          & \multirow{9}[0]{*}{From KDE} & \multirow{2}[0]{*}{From KDE} \\
		$ q $                         &       1.7203      &             Uniform(1, 2)           &               &                               \\
		$ D_L $ [Mpc]                 &        1000       &        UniformRadius(100, 2000)     &               &      Fixed: 998.3             \\
		$ \iota $ [rad]               &       0.7854      &            Sine(0, $\pi$)           &               &      Fixed: 0.7863            \\
		$ \Delta t_c $ [s]            &       0           &        Uniform($-0.001$, 0.001)     &               &      Fixed: $-0.00002$        \\
		$ \phi_c $ [rad]              &       4.2759      &         Uniform(0, $2\pi$)          &               &      Fixed: 4.2730            \\
		$ \psi $ [rad]                &       1.0472      &           Uniform(0, $\pi$)         &               &      Fixed: 1.0470            \\
		$ \alpha $ (RA) [rad]         &       2.9019      &   \multirow{2}[0]{*}{UniformSky}    &               &      Fixed: 2.9019            \\
		$ \delta $ (DEC) [rad]        &     $-0.4704$     &                                     &               &      Fixed: $-0.4713$         \\
		$ e_{0.01\mathrm{Hz}} $       &       0.1         &                 -                   & Uniform(0.09, 0.11) & Uniform(0.09, 0.11)  \\
		\hline
		\hline
	\end{tabular}%
	\label{prior}%
\end{table*}                    

Following the frequency-domain response of Marsat et al. \cite{Marsat2018,Marsat2021}, a single link response reads
\begin{equation}\label{ysr}
	{{\tilde y}_{sr}} = \sum\limits_{\ell ,m} {\mathcal{T}_{sr}^{\ell m}(f){{\tilde h}_{\ell m}}},
\end{equation}
with transfer function
\begin{equation}\label{tsr}
	\mathcal{T}_{sr}^{\ell m}(f)=G_{sr}^{\ell m}\left(f, t_{f}^{\ell m}\right),
\end{equation}
\begin{equation}\label{gsr}
	\begin{aligned}
		G_{sr}^{\ell m}(f, t) & = \frac{{\rm i} \pi f L}{2} \operatorname{sinc}\left[\pi f L\left(1-k \cdot n_{l}\right)\right] \\
		& \cdot \exp \left[{\rm i} \pi f\left(L+k \cdot\left(p_{r}+p_{s}\right)\right)\right] n_{l} \cdot P_{\ell m} \cdot n_{l},
	\end{aligned}
\end{equation}
where $k$ is the wave vector of \acp{GW}, $ n_l $ is the link unit vector from spacecraft $ s $ to $ r $, and $ p_s, p_r $ are the positions of the spacecrafts at the same time $ t $.
In Eq.(\ref{tsr}),
\begin{equation}\label{tflm}
	t_f^{\ell m} = \frac{1}{{2\pi }}\frac{{d{\Psi _{\ell m}}}}{{df}},
\end{equation}
where $ \Psi _{\ell m} $ stands for the slowly varying phase of a specific harmonic $ (\ell,m) $:
\begin{equation}
	{{\tilde h}_{\ell m}}(f) = {A_{\ell m}}{e^{{\rm i}{\Psi _{\ell m}}}}.
\end{equation}

Eq.(\ref{tflm}) describes the effective time-frequency correspondence for each spin-weighted spherical harmonic, which is a generalization of the \ac{SPA}.
Although \texttt{EccentricFD} only has the dominant spin-weighted spherical harmonic $ (\ell, m)=(2, 2) $, eccentric radiation appears as harmonics $j$ with distinct time-frequency correspondences, i.e.
\begin{equation}
	t_f^j = \frac{1}{{2\pi }}\frac{{d{\Psi _j}}}{{df}}.
\end{equation}
Since $ \xi _j^{ + , \times } $ is entangled in the waveform and is a function of the evolving eccentricity, we could not simply write $ H $ as in Eq.(\ref{hlm}).
Instead, we go back to Eq.(\ref{Hwf}):
\begin{equation}
	\begin{aligned}
		H &=\sum\limits_{j} h_j,\\
		{\tilde h}_j &={{P_ + }\tilde h_j^ +  + {P_ \times }\tilde h_j^ \times },
	\end{aligned}
\end{equation}
and then generalize Eqs. \eqref{ysr}-\eqref{gsr} by treating each eccentric harmonic separately:
\begin{equation}
	{{\tilde y}_{sr}} = \sum\limits_j {\mathcal{T}_{sr}^j(f):{\tilde h}_j},
\end{equation}
\begin{equation}
	\mathcal{T}_{sr}^j(f) = {G_{sr}}\left( {f,t_f^j} \right),
\end{equation}
\begin{equation}\label{gsr_e}
	\begin{aligned}
		{G_{sr}}(f,t) &= \frac{{{\rm i}\pi fL}}{2}\operatorname{sinc} \left[ {\pi fL\left( {1 - k \cdot {n_l}} \right)} \right] \\
		&\cdot \exp \left[ {{\rm i}\pi f\left( {L + k \cdot \left( {{p_r} + {p_s}} \right)} \right)} \right]{n_l} \otimes {n_l},
	\end{aligned}
\end{equation}
where : denotes the double contraction, $A:B=\sum_{pq}A_{pq}B_{pq}$.
This edited form enables frequency-domain \ac{TDI} responses to be applied consistently to eccentric harmonics.
The waveform and \ac{TDI} response described above are implemented in \texttt{GWSpace}~\cite{Li2025} and are used for inference via the \texttt{PyCBC} waveform plugin.

\begin{figure*}
	\centering
	\includegraphics[width=\linewidth]{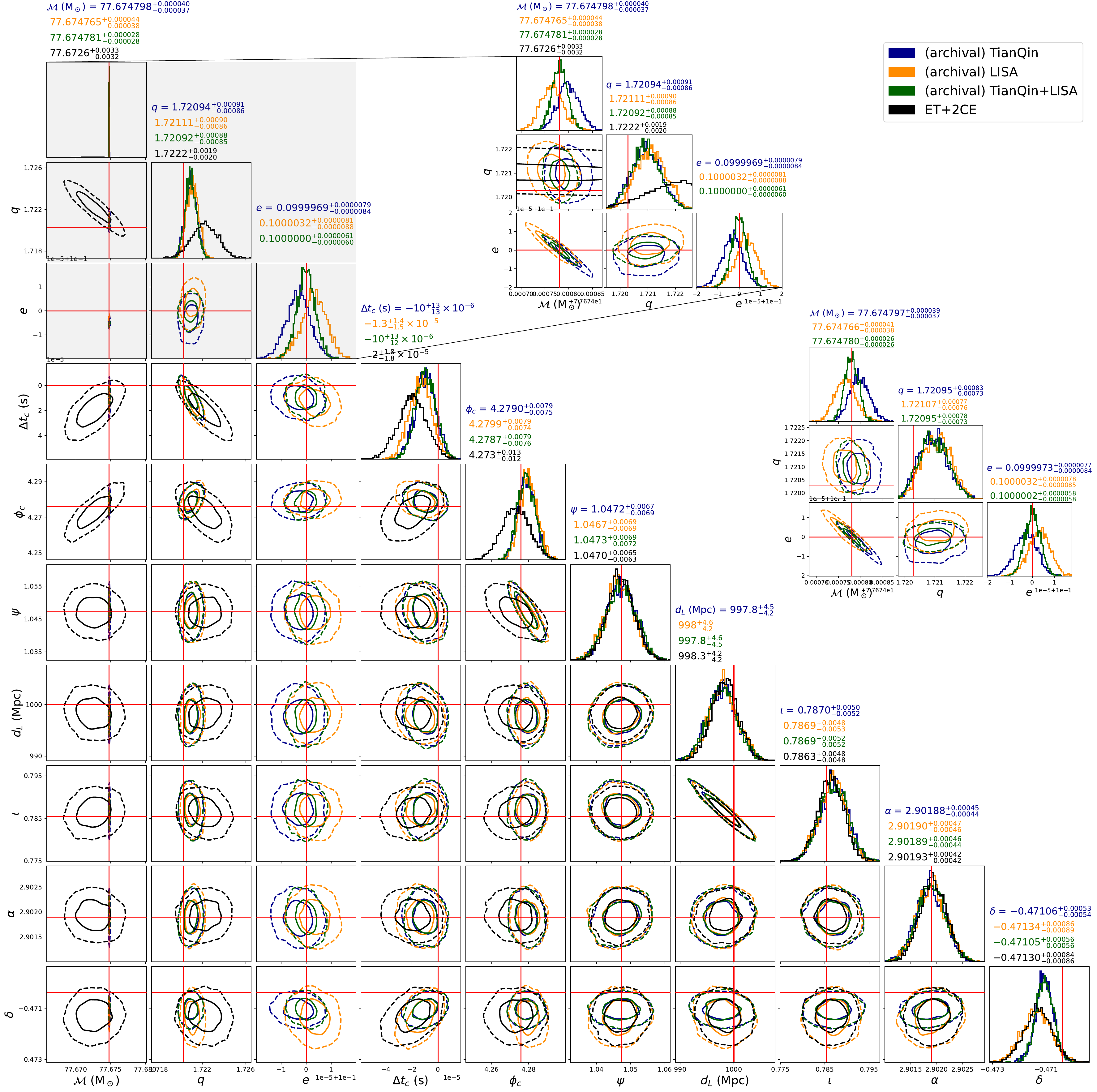}
	\caption{Posterior distributions for a GW190521-like source.
		The main panel shows the full ten-dimensional posterior distribution.
		Black curves denote results obtained with the next-generation ground-based detector network (ET+2CE), using the injection and priors listed in Table \ref{prior}.
		Blue, yellow, and green curves correspond to space-based analyses with TianQin, LISA, and TianQin+LISA, respectively, employing the ground-informed priors summarized in Table \ref{prior} (Space:10D).
		The upper-right inset presents a zoomed-in view of the chirp mass, mass ratio and eccentricity, highlighting the detailed structure of the posterior peak.
		The lower-right inset shows a posterior distribution similar to the main one but with all parameters fixed at the ground-based best-fit values except the mass parameters and eccentricity (i.e., using the ground-informed priors in Table \ref{prior}, Space:3D).
		The red line marks the injected (``true'') values.
		Two-dimensional contours enclose 50\% and 90\% of the posterior.}
	\label{pe}
\end{figure*}

\section{Capability of the ground-detector network\label{sec3}}

To determine the parameter space that must be explored in the archival-search scenario, we first need to assess how it is constrained by next-generation ground detectors.
Throughout this work, we neglect possible signal overlap and analyze data consisting of stationary Gaussian noise plus a single signal for both ground- and space-based detectors.
We consider a GW190521-like\cite{Abbott2024} injection in the ET+2CE network, using the injected values and priors in Table~\ref{prior}.
The resulting posteriors are summarized in Fig.~\ref{pe}.
Because of the improved sensitivity relative to current detectors, the network \ac{SNR} is $\sim 1.5\times10^3$ and all parameters are tightly constrained.
Comparing with space-detector capability estimates~\cite{Sesana2016,Liu2020,Lyu2023}, we find that for quasi-circular sources the parameters that space detectors can measure more precisely than the ground network are the mass parameters, consistent with previous multiband studies~\cite{Ewing2021,Klein2022,Wang2024}.

Motivated by this, we set the prior of the space-based parameter estimation from the ground-based posterior.
To convert discrete posterior samples into a continuous probability distribution, the prior is constructed via a kernel density estimate (KDE) of the corresponding ground posterior, a widely used approach in this context. In our case, the ground posteriors are unimodal and smooth, so a Gaussian-kernel KDE provides a sufficiently accurate reconstruction.
In future applications, our simulation-based ground posteriors should be replaced by the published posteriors of the relevant events.
Using released ground posteriors as priors--rather than performing a fully joint multiband parameter estimation--is the practical choice for archival searches and yields a lower-bound estimate of the parameter-space volume to be covered, and thus of the computational cost.

\section{Eccentricity in space-based Parameter Estimation\label{sec4}}

Based on the setup described above, we perform space-based parameter estimation for TianQin, LISA, and their network (Fig. \ref{pe}).
We also present results in which only the component masses and the eccentricity are allowed to vary (Fig. \ref{pe}, lower-right inset).
All parameters except $(\mathcal{M}, q, e_{0.01\mathrm{Hz}})$ are fixed to the median values of the ground-based posteriors rather than the injected ones, while these three parameters remain well constrained in all space-based analyses.
We further verified that, in the space-band analysis, adopting the full ground posterior KDE as the prior for all parameters except eccentricity yields only minor improvements for the non-mass parameters, producing no significant differences in $(\mathcal{M}, q, e_{0.01\mathrm{Hz}})$ between the reduced three-dimensional and full-dimensional parameter-estimation runs.
Notably, the chirp mass $\mathcal{M}$ is measured about two orders of magnitude more precisely than in the ground-only case (from $\mathcal{O}(10^{-3})$ to $\mathcal{O}(10^{-5})M_\odot$).
Although space data alone are not more informative about $q$ than the ground network, the strong constraint on $\mathcal{M}$ compresses the $\mathcal{M}$-$q$ degeneracy, yielding a $q$ marginal posterior that is less than half as wide as the ground-only result.
A similar improvement is observed for the coalescence time and phase, which are partially degenerate with the two mass parameters.
With $\mathcal{M}$ tightly constrained, the dominant residual degeneracy is between $\mathcal{M}$ and the orbital eccentricity $e_{0.01\mathrm{Hz}}$, as anticipated by previous studies~\cite{Lenon2020,OShea2023,Favata2022}.
In our GW190521-like configuration, the eccentricity is constrained to $\mathcal{O}(10^{-5})$ around the injected value $e_{0.01\mathrm{Hz}}=0.1$.
Quantitatively, the precisions across parameters are consistent with previous estimates of space-band and multiband capabilities~\cite{Liu2020,Toubiana2022, Wu2025}.

We validated the heterodyned likelihood for eccentric waveforms.
It accurately reproduces the peak location and local curvature of the Gaussian likelihood, and the Jensen-Shannon divergence (JSD) between the two posteriors satisfies the adopted criterion (see Appendix~\ref{app} for details).
Using the heterodyned likelihood can reduce computational cost from $\mathcal{O}(10^3)-\mathcal{O}(10^4)$ core-hours to $\mathcal{O}(10)-\mathcal{O}(10^2)$ core-hours and substantially lower the memory usage.

We highlight that we adopted a GW190521-like configuration with AET-channel network \ac{SNR} $\sim 7$ for each of TianQin and LISA, and $\sim 10$ for TianQin+LISA.
Parameter precision is largely SNR-driven, the standalone TianQin and LISA results are similar.
We emphasize that TianQin and LISA \acp{SNR} are not generally comparable \cite{Chen2025}.
The \ac{SNR} depends strongly on source parameters and sky location.
Smaller chirp masses shift signals to higher frequencies, which will be favorable for TianQin.
Additionally, TianQin's fixed orbital plane (aligned with J0806) yields strongly sky-dependent sensitivity (see Appendix~\ref{app_snr}).
This directional response also explains why TianQin provides slightly tighter constraints on the declination angle~$\delta$ for certain source locations, as seen in Fig.~\ref{pe}.
A joint TianQin+LISA observation leverages complementary sky responses, so that regions disfavored by one constellation are compensated by the other, enhancing the \ac{SNR} and robustness of parameter estimation.

Detection thresholds quoted for untargeted searches are often $\rho_{\rm thr}\gtrsim8$, but in our archival search setting, one can confidently claim a detection at \ac{SNR} $\sim7$ .
The key point is that ground detections pass informative priors to the space analysis.
We can estimate the \ac{SNR} threshold via the size of matched-filtering template bank~\cite{Moore2019a}
\begin{equation}\label{thr}
	\rho_{\rm thr}=\sqrt{-2\ln\left(\frac{\mathrm{FAP}}{A\times N_T}\right)},
\end{equation}
where $\mathrm{FAP}$ is the false-alarm probability and $A\times N_T$ is the number of effectively independent templates.
By setting $A=1$ we can perform direct comparisons with previous studies \cite{Moore2019a,Ewing2021}.
Ground-based detectors typically require a detection threshold $ \mathrm{FAP}\sim10^{-3} $.
Using the bank size estimated in our previous study~\cite{Wang2024} for an archival search over chirp mass and eccentricity, $N_T\sim\mathcal{O}(10^{8})$, it gives $\rho_{\rm thr}\approx7.12$.
In the multiband context, one may adopt a larger $\mathrm{FAP}$ for space because credibility is already established by the ground detection.
Then candidates with network \ac{SNR} $\sim$7 (and even below) can be distinguished in multiband searches, which can boost the overall detection number~\cite{Wu2025}.

\section{Summary and discussion\label{sec5}}
Orbital eccentricity in \acp{sBBH} is a key tracer of formation channels, yet extracting it from \ac{GW} data remains challenging.
We implemented a Bayesian inference pipeline for \ac{sBBH} signals found through archival searches that, for the first time in this context, includes eccentricity.
By conditioning on next-generation ground-network information, our framework enables space-band inference in noisy data and makes eccentric \ac{sBBH} analyses practically accessible.
In particular, the space-based analysis yields substantially tighter constraints on the chirp mass and robust, high-precision constraints on the orbital eccentricity.

Using a GW190521-like source as an example, we employ ground posteriors to define space-band priors: a KDE prior in $(\mathcal{M},q)$ while fixing the remaining parameters to the ground posterior medians.
This practical archival workflow avoids reanalyzing ground data yet captures the constraints from the ground network.
In this setup, space observations sharpen the chirp mass from $\mathcal{O}(10^{-3})$ to $\mathcal{O}(10^{-5})M_\odot$ and constrain the eccentricity to $\mathcal{O}(10^{-5})$ around the injected value $e_{0.01\mathrm{Hz}}=0.1$.
The degeneracy between eccentricity and chirp mass is consistent with previous studies.
We also show that archival searches can adopt lower space-band \ac{SNR} thresholds than blind searches: ground-informed priors reduce the effective template-bank size and thus the threshold, from $\rho_{\rm thr}\sim15$ to $\sim7.12$ under conservative assumptions.
Consequently, candidates with network \ac{SNR} $\sim7$ (or even below) can be distinguished and tightly constrained in practice, enabling more low-SNR space-band counterparts and increasing the number of multiband detections.
This strengthens \ac{sBBH} population studies (especially eccentricity-informed formation mechanisms) as well as downstream tests of environmental effects and possible deviations from general relativity.
The newly released GWTC-4 catalog further highlights the promise of the space band \cite{Abac2025a,Abac2025b}.
The enlarged catalog strengthens population analyses and, notably, includes the most massive event so far, GW231123\cite{Abac2025}.
Because higher-mass \acp{sBBH} shift power to lower frequencies, such systems will be particularly favorable for space-based detectors.
Furthermore, joint observations can take advantage of complementary sky responses of TianQin and LISA, delivering a higher network sensitivity.

we adopt the heterodyned likelihood for calculation, which reduces computational cost from $\mathcal{O}(10^3)-\mathcal{O}(10^4)$ to $\mathcal{O}(10)-\mathcal{O}(10^2)$ core-hours and lowers memory usage, making inference with eccentricity broadly accessible.
For higher eccentricities, tailored or more general acceleration strategies will be valuable so that additional parameters can be accommodated in archival searches.
For example, we adopt an optimistic ET+2CE ground network, realistic duty cycles will degrade constraints on parameters (e.g., sky localization) when not all detectors are operating, and this should be accounted for in future applications.
Duty-cycle losses also affect the space band\cite{Luo2016,Seoane2022}, but for the one-year observation considered here our conclusions remain robust: the order-of-magnitude precision gains are essentially unchanged, though duty cycle should be modeled in analyses of real data.
We use \texttt{EccentricFD}, a non-spinning eccentric waveform, and neglect spins in the ground band for consistency.
Spin effects during inspiral are expected to be subdominant here and tightly constrained by ground data in practice, but waveform accuracy remains critical to avoid systematics in parameter inference and in subsequent tests \cite{RomeroShaw2021,RomeroShaw2023,Gupta2025}.

The data supporting the findings of this study will be made publicly available at \url{https://github.com/HumphreyWang/eccentric_bbh_multiband_public}.

\section{Acknowledgments}
We are grateful to Shichao Wu and Alexander Nitz for the helpful discussions.
This work has been supported by the National Key Research and Development Program of China (Grant No. 2023YFC2206700).
Bayesian inference computations were primarily performed on the SCIAMA HPC cluster, supported by the Institute of Cosmology and Gravitation, University of Portsmouth.
H.W. acknowledges financial support from the China Scholarship Council (CSC) for the research visit to the University of Portsmouth.
M.J.W. and I.H. acknowledge support from UKSA through Grants No. ST/X002225/1 and No. ST/Y004876/1.
Y.H. is supported by the Natural Science Foundation of China (Grants  No.  12173104), the science research grants from the China Manned Space Project (CMS-CSST-2025-A13), Fundamental Research Funds for the Central Universities, Sun Yat-sen University, and the 111 Project (Grant No.B20062).

\bibliographystyle{unsrtnat}
\bibliography{ref}

\appendix
\section{Validity of heterodyned likelihood\label{app}}

The heterodyned likelihood assumes that the waveform amplitude and phase vary slowly with frequency.
This assumption can be challenged by eccentric signals, which contain multiple harmonic components.
To assess validity, we fix all parameters except the eccentricity and evaluate the log-likelihood ratio (LLR) on a uniform grid in $e_{0.01\mathrm{Hz}}$ using both the standard Gaussian likelihood and the heterodyned likelihood. We define
\begin{equation}
	\mathrm{LLR}(\lambda) \equiv \log\frac{P(D|\lambda^\mu)}{P(D|n)} 
	= (h|D) - \frac{1}{2}(h|h).
\end{equation}

\begin{figure}[ht]
	\centering
	\includegraphics[width=\linewidth]{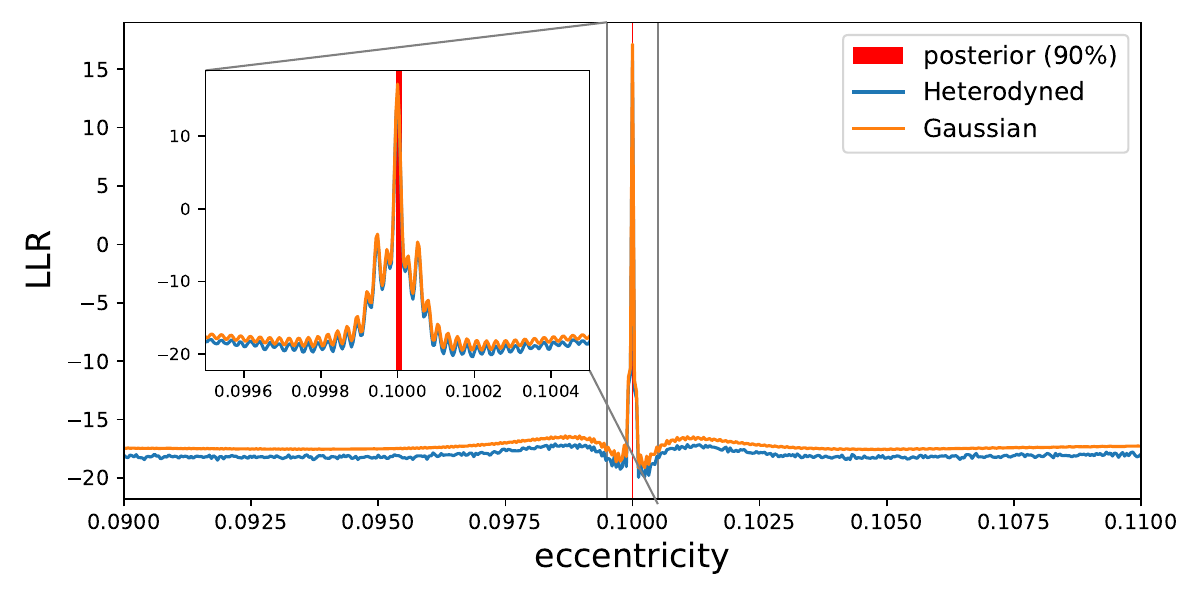}
	\caption{Comparison between the Gaussian likelihood and the heterodyned likelihood for a LISA analysis at true value $e_{0.01\mathrm{Hz}}=0.1$.
	The orange (blue) curve shows the Gaussian (heterodyned) log-likelihood ratio.
	The red band marks the 90\% credible interval of the one-dimensional posterior for $e_{0.01\mathrm{Hz}}$ from Fig.~\ref{pe}.
	To better highlight differences between the likelihood models, here we assume zero-noise data.}
	\label{loglr}
\end{figure}

At $e_{0.01\mathrm{Hz}}=0.1$, both methods recover the same maximum and exhibit nearly identical curvature around the peak (Fig.~\ref{loglr}).
Small and smooth fluctuations in the heterodyned likelihood curve arise from accelerated waveform evaluation, and a global vertical offset reflects a constant difference in the noise term.
Neither affects parameter inference.
More importantly, the red band (the 90\% credible interval from Fig.~\ref{pe}) is much narrower than the scale over which the two curves significantly diverge and is centered on the maximum.

We also found that, for smaller true eccentricity, the two curves become indistinguishable.
This indicates that the mild discrepancies seen at $e_{0.01\mathrm{Hz}}=0.1$ attributes to the increasing importance of eccentric harmonics, which weakens the ``slowly varying'' assumption.
Nevertheless, within the narrow prior ranges relevant for archival searches, the heterodyned approximation accurately reproduces the peak location and local curvature of the likelihood, yielding posteriors effectively identical to those from the full Gaussian likelihood. 

To further validate this, we compute the Jensen-Shannon divergence (JSD) between the two posteriors under different configurations and adopt a conservative criterion of Ref.~\cite{Ashton2021}:
\begin{equation}
	\max \mathrm{JSD}\le\frac{10}{n^{\mathrm{eff}}_{\mathrm{samples}}} .
\end{equation}
With $n^{\mathrm{eff}}_{\mathrm{samples}}=1000$, the threshold is $0.01$, and our maximum JSD is $\approx 0.0095<0.01$.
Hence, for $e_{0.01\mathrm{Hz}}\lesssim 0.1$, the heterodyned likelihood is a good approximation for our use case.

\section{Sky-orientation dependence of \ac{SNR} for TianQin/LISA\label{app_snr}}
\begin{figure}
	\centering
	\includegraphics[width=\linewidth]{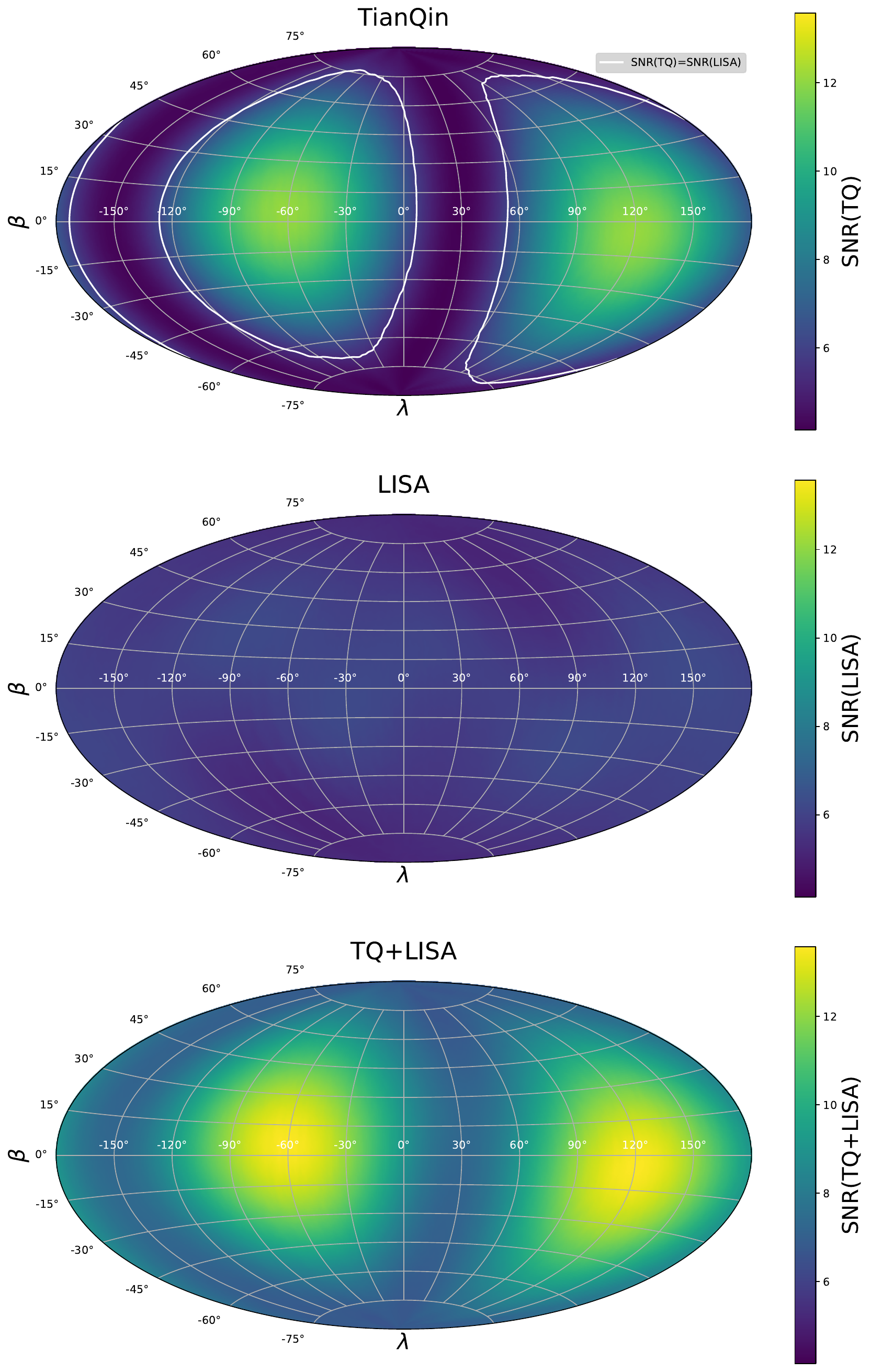}
	\caption{Sky maps of the network \ac{SNR} for a GW190521-like source observed with TianQin (top), LISA (middle), and the joint TianQin+LISA network (bottom).
	The white contour on the TianQin panel marks sky locations where $\mathrm{SNR}_{\rm TQ}=\mathrm{SNR}_{\rm LISA}$ for this setup. }
	\label{fig:gw190521asnr}
\end{figure}

We sample $401\times201$ points uniformly in ecliptic longitude and latitude, fix all other parameters to the values in Table~\ref{prior}, and compute the total \ac{TDI} \ac{SNR}.
The resulting sky maps for the different detectors are shown in Fig.~\ref{fig:gw190521asnr}, with a common color scale for direct comparison.

TianQin's constellation plane is fixed and points toward the verification binary J0806 at $(\lambda,\beta)=(120.5^\circ,-4.7^\circ)$.
Sources near that direction (and its antipode) achieve the highest \ac{SNR}, whereas sources near the great circle defined by the intersection of the TianQin plane with the celestial sphere yield the lowest \ac{SNR}, producing strong sky anisotropy.
In contrast, LISA's constellation plane is inclined by $\sim60^\circ$ to the ecliptic and precesses over the year, averaging the response and yielding a much weaker sky dependence.
To highlight the contrast, we overlay on the TianQin map the zero-difference contour of $\mathrm{SNR}_{\rm TQ}-\mathrm{SNR}_{\rm LISA}$.
Note that this contour is specific to the source and parameter choices in Table~\ref{prior} and will shift with parameter sets.

For the joint TianQin+LISA network, the minimum network \ac{SNR} across the sky reaching $\gtrsim 6.5$ in this configuration.
This mitigates TianQin's anisotropy, improves sky coverage and robustness to source location, and yields tighter parameter constraints.

\end{document}